\documentstyle[12pt]{article}
\newcommand{\bce}{\begin{center}}
\newcommand{\ece}{\end{center}}
\newcommand{\beq}{\begin{equation}}
\newcommand{\eeq}{\end{equation}}
\newcommand{\bea}{\vspace{0.25cm}\begin{eqnarray}}
\newcommand{\eea}{\end{eqnarray}}

\newcommand{\ba}{\begin{array}}
\newcommand{\ea}{\end{array}}

\newcommand{\doublespace}{
    \renewcommand{\baselinestretch}{1.6}\large\normalsize}

\def\lsim{\mathrel{\rlap{\lower4pt\hbox{\hskip1pt$\sim$}}
    \raise1pt\hbox{$<$}}}         
\def\gsim{\mathrel{\rlap{\lower4pt\hbox{\hskip1pt$\sim$}}
    \raise1pt\hbox{$>$}}}         

\def\Pom{{\bf I\!P}}

\def\lsim{\mathrel{\rlap{\lower4pt\hbox{\hskip1pt$\sim$}}
    \raise1pt\hbox{$<$}}}         
\def\gsim{\mathrel{\rlap{\lower4pt\hbox{\hskip1pt$\sim$}}
    \raise1pt\hbox{$>$}}}         

\def\Pom{{\bf I\!P}}

  \advance\oddsidemargin  by -1in
  \advance\evensidemargin by -1in
  \advance\topmargin      by -0.5in
\def\lsim{\mathrel{\rlap{\lower4pt\hbox{\hskip1pt$\sim$}}
    \raise1pt\hbox{$<$}}}         
\def\gsim{\mathrel{\rlap{\lower4pt\hbox{\hskip1pt$\sim$}}
    \raise1pt\hbox{$>$}}}         

\def\Pom{{\bf I\!P}}
\def\beq{\begin{equation}}
\def\endeq{\end{equation}}
\def\arr{\begin{eqnarray}}
\def\endarr{\end{eqnarray}}
\makeindex

\doublespace

\begin{document}

\phantom{.}{\large \bf \hspace{9.4cm} ITEP-74-94\\
\phantom{.}\hspace{10.2cm}10 October 1994\\ }

\begin{center}
{\bf\sl \huge The direct calculation of the slope of    \\
the QCD pomeron's trajectory}
\vspace{0.4cm}\\
{\bf \large
N.N.~Nikolaev$^{a,b}$, B.G.~Zakharov$^{b}$ and V.R.Zoller$^{c}$
\bigskip\\}
{\it
$^{a}$IKP(Theorie), KFA J{\"u}lich, 5170 J{\"u}lich, Germany
\medskip\\
$^{b}$L. D. Landau Institute for Theoretical Physics, GSP-1,
117940, \\
ul. Kosygina 2, Moscow 117334, Russia.\medskip\\
$^{c}$ Institute for Theoretical and Experimental Physics,\\
Bolshaya Cheremushkinskaya 25, 117259 Moscow, Russia.
\vspace{1.0cm}\\ }
{\Large
Abstract}\\
\end{center}
We demonstrate that the diffraction slope of the generalized
BFKL pomeron amplitude has the conventional Regge growth $B(s)
= B(0) + 2\alpha_{\Pom}'\log(s)$. This proves that the
generalized BFKL pomeron is described by the moving $j$-plane
singularity. We  give an estimate for the slope $\alpha_{\Pom}'$
in terms of the correlation radius for the perturbative gluons.

\pagebreak




\section{Introduction}
Whether the QCD pomeron is described by the fixed or moving
singularity in the complex $j$-plane, remains one of the
topical issues. The purpose of this communication is a proof
that the generalized BFKL pomeron [1-4] is a moving cut.
We present the first direct calculation of the slope $\alpha_{\Pom}'$
for the pomeron trajectory.

The early works on the BFKL (Balitskii-Fadin-Kuraev-Lipatov [5])
pomeron focused on the idealized
scaling regime with fixed strong coupling
$\alpha_{S}=const$ and infinite gluon correlation radius $R_{c}$.
In this regime, the BFKL pomeron is described
by a fixed cut in the complex angular momentum plane $-\infty < j
\leq \alpha_{\Pom}(0)= 1+\Delta_{\Pom}$. However, because of the
diffusion property of the Green function of the scaling BFKL
equation [5], the scaling regime is not self-consistent.
Recently, there was much progress in understanding the BFKL
pomeron in the framework of the dipole-cross section representation
introduced in [6]. In previous papers of ours [1-4] we derived
the generalized BFKL equation for the dipole cross section in a
realistic model with the running (and freezing) strong coupling
$\alpha_{S}(r)$ and with the finite corelation radius $R_{c}$ of
the perturbative gluons. While the property of the cut in the
$j$-plane is retained, we found a profound impact of the running
$\alpha_{S}(r)$ and of the finite $R_{c}$ on the spectrum and
solutions of our generalized BFKL equation. The crucial
observation is
that the intercept $\Delta_{\Pom}$ and the asymptotic behaviour
of the dipole cross section are controlled by interactions at the
dipole size $r\sim R_{c}$. Also, we noticed that the recovery of
the conventional multiperipheral pattern is likely at asymptotic energies,
which suggests the Regge growth of the diffraction cone.
In this paper we confirm the latter observation and show that
indeed the pomeron trajectory has the finite slope
$\alpha_{\Pom}'\propto R_{c}^{2}$.

The starting point of our analysis is the generalization of
our BFKL equation [1-4] to the profile function of the dipole
cross section $\Gamma(r,\vec{b})$. Defining the impact parameter
$\vec{b}$ with respect to the center of the parent $q$-$\bar{q}$
dipole, and repeating the derivation [1-4], we obtain
\arr
{\partial \Gamma(\xi,r,\vec{b})\over \partial \xi} =
{\cal K}\otimes \Gamma(\xi,r,\vec{b})~~~~~~~~~~~~~~~~~~~~~~~~~~
 \nonumber\\
={3 \over 8\pi^{3}} \int d^{2}\vec{\rho}_{1}\,\,
\mu_{G}^{2}
\left|g_{S}(R_{1})
K_{1}(\mu_{G}\rho_{1}){\vec{\rho}_{1}\over \rho_{1}}
-g_{S}(R_{2})
K_{1}(\mu_{G}\rho_{2}){\vec{\rho}_{2} \over \rho_{2}}\right|^{2}
\nonumber\\
\times
[\Gamma(\xi,\rho_{2},\vec{b}+{1\over 2}\vec{\rho}_{1}) +
\Gamma(\xi,\rho_{1},\vec{b}+{1\over 2}\vec{\rho}_{2}) -
\Gamma(\xi,r,\vec{b})]  \, .
\label{eq:1}
\endarr
in which $\vec{\rho}_{2}=\vec{\rho}_{1}-\vec{r}$, the arguments
of the running QCD charge $g_{S}(r)=\sqrt{4\pi \alpha_{S}(r)}$
are $R_{i}={\rm min}\{r,\rho_{i}\}$, $K_{1}(x)$ is the generalized
Bessel function and  $R_{c}=1/\mu_{G}$ is the correlation radius
for the perturbative gluons. Here we use the standard
definition of the profile function when
$A(s,t)=2is\int d^2\vec{b}\exp(-i\vec{q}\vec{b})\Gamma(\vec{b})$
and the dipole cross section equals
$\sigma(\xi,r)=2\int d^2\vec b\, \Gamma(\xi,r,\vec{b})$.
Hereafter we shall discuss the reduction of (\ref{eq:2}) to the
equation for the diffraction slope
$B(\xi,r)= {1\over 2}\langle \vec{b}\,^{2}\rangle =
\lambda(\xi,r)/\sigma(\xi,r)$,
where
$
\lambda(\xi,r)=\int d^2\vec{b}~  \vec{b}\,^2~\Gamma(\xi,r,\vec{b})
$.
Evidently, the diffraction slope for the dipole of size $r$
contains the purely geometrical contribution ${1\over 8}r^{2}$
which comes from the elastic form factor of the dipole. Then, it is
more convenient to consider
$
\eta(\xi,r)=\lambda(\xi,r)-{1\over 8}r^{2}\sigma(\xi,r) \, ,
$
the equation for which takes the form
\arr
{\partial \eta(\xi,r) \over \partial \xi} =
{3 \over 8\pi^{3}} \int d^{2}\vec{\rho}_{1}\,\,
\mu_{G}^{2}
\left|g_{S}(R_{1})
K_{1}(\mu_{G}\rho_{1}){\vec{\rho}_{1}\over \rho_{1}}
-g_{S}(R_{2})
K_{1}(\mu_{G}\rho_{2}){\vec{\rho}_{2} \over \rho_{2}}\right|^{2}
\nonumber\\
\times\left\{\eta(\xi,\rho_{1})+
\eta(\xi,\rho_{2})-\eta(\xi,r)+
{1\over 8}(\rho_{1}^{2}+\rho_{2}^{2}-r^{2})[
\sigma(\rho_2,\xi) + \sigma(\rho_1,\xi)]\right\}\nonumber\\
={\cal K}\otimes \eta(\xi,r) +\beta(\xi,r)
 \, ,
\label{eq:2}
\endarr
where the inhomegeneous term equals
\arr
\beta(\xi,r) = {\cal L}\otimes \sigma(\xi,r) ~~~~~~~~~~~~~~~~~~~~~~~~~~
 \nonumber\\
={3 \over 64\pi^{3}} \int d^{2}\vec{\rho}_{1}\,\,
\mu_{G}^{2}
\left|g_{S}(R_{1})
K_{1}(\mu_{G}\rho_{1}){\vec{\rho}_{1}\over \rho_{1}}
-g_{S}(R_{2})
K_{1}(\mu_{G}\rho_{2}){\vec{\rho}_{2} \over \rho_{2}}
\right|^{2} \nonumber\\
(\rho_{1}^{2}+\rho_{2}^{2}-r^{2})[
\sigma(\rho_2,\xi) + \sigma(\rho_1,\xi)]
\, \, .
\label{eq:3}
\endarr
Here the crucial point is that the homogeneous Eq.~(\ref{eq:2})
is precisely our generalized BFKL equation for the
dipole cross section
\beq
{d\sigma(\xi,r)\over d\xi} = {\cal K}\otimes
\sigma(\xi,r)\, ,
\label{eq:4}
\endeq
which enables us to prove on a very generic grounds that
$\alpha_{\Pom}'={1\over 2}dB(\xi,r)/d\xi \neq 0$.

The proof goes as follows: In [3,4] we have shown that the
generalized BFKL operator ${\cal K}$ has the continuous spectrum,
which corresponds to the cut in the $j$-plane.
Let $-\infty <\nu <\infty$ be the "wavenumber" which labels
eigenfunctions $E(\nu,r)\exp[\Delta(\nu)\xi]$ of Eq.~(\ref{eq:4})
with eigenvalue $\Delta(\nu)$. For the guidance,
in the scaling limit of $\alpha_{S} =const$ and $R_{c}\rightarrow
\infty$ $ E(\nu,r)=r\exp[i\nu\log(r^{2})]=
\sigma_{\Pom}(r)\exp[i\nu\log(r^{2})]
$
with the orthogonality condition [3-5]
\beq
\delta(\nu-\mu)=  {1 \over 2\pi}
\int {d\log(r^2) \over  [\sigma_{\Pom}(r)]^{2} }
E^{*}(\nu,r)E(\mu,r) \, ,
\label{eq:5}
\endeq
and $\nu$ is indeed the wavenumber of plane waves in the
$\log(r^{2})$ space. Properties of eigenfunctions $E(\nu,r)$ in
the case of the running $\alpha_{S}(r)$ and the finite $R_{c}$ are
discussed in [3,4,7].

Now we proceed with solution of the inhomogeneous equation
(\ref{eq:2}). If
$
G(\nu,r)={\cal L}\otimes E(\nu,r)
= \int d\omega g(\nu,\omega)
E(\omega,r)\, ,
$
then the inhomogeneous term (\ref{eq:3}) can be written as
\beq
\beta(\xi,r) = {\cal L}\otimes \sigma(\xi,r) =
\int d\nu E(\nu,r)\int d\omega f(\omega) g(\omega,\nu)
\exp[\Delta(\omega)\xi] \, .
\label{eq:6}
\endeq
We search for a solution of the form
$
\eta(\xi,r)= \int d\nu \tau(\xi,\nu)
E(\nu,r) \exp[\Delta(\nu)\xi]\,.
$
Making use of the property of eigenfunctions
${\cal K\otimes}E(\nu,r)=\Delta(\nu)E(\nu,r)$ we
find
\beq
{\partial \tau(\xi,\nu) \over \partial\xi}
\exp[\Delta(\nu)\xi] =
\int d\omega f(\omega)
g(\omega,\nu)
\exp[\Delta(\omega)\xi]
\label{eq:7}
\endeq
and
\arr
\eta(\xi,r)=
\int d\nu
\tau(\xi=0,\nu)E(\nu,r) \exp[\Delta(\nu)\xi]  \nonumber\\
+
\int_{0}^{\xi} d\xi'
\int d\nu E(\nu,r) \exp[\Delta(\nu)(\xi-\xi')]
\int d\omega f(\omega)
g(\omega,\nu)
\exp[\Delta(\omega)\xi'] \, .
\label{eq:8}
\endarr
Here $\tau(\xi=0,\nu)$ describes a solution of the homogeneous
equation (\ref{eq:2}) and is determined by the initial condition
$\eta(\xi=0,r)$.

The singularity structure of $g(\omega,\nu)$ can be found
considering the large-$r$ behaviour of $G(\nu,r)={\cal L}\otimes
E(\nu,r)$. Because of the exponential decrease of the Bessel
function $K_{1}(x)\propto \exp(-x)$, the integration in (\ref{eq:3})
will be dominated by the two contributions from $\rho_{1} \lsim
R_{c},~\rho_{2}\approx r$ and $\rho_{2}\lsim R_{c},~\rho_{1}
\approx r$. For the sake of definiteness, consider the former case.
Notice that, in this regime, $(\rho_{1}^{2}+\rho_{2}^{2}-r^{2})
\approx \rho_{1}^{2}$ and $E(\nu,\rho_{2})\approx E(\nu,r)$, which
gives the contribution of the form $2G_{1}E(\nu,r)$ to ${\cal L}
\otimes E(\nu,r)$. Evidently, it corresponds to the singular term
$g_{1}(\omega,\nu)=2G_{1}\delta(\omega-\nu)$. The contribution
from the term $\propto \rho_{1}^{2}E(\nu,\rho_{1})$ to
${\cal L}\otimes E(\nu,r)$ does not depend on $r$ at large $r$
and corresponds to $g_{2}(\omega,\nu)=
G_{2}(\nu)\delta(\omega)$. Apart from these singular terms,
$g(\omega,\nu)$ will also have the smooth component $g_{3}(\omega,
\nu)$.

Evidently, the $2G_{1}\delta(\omega-\nu)$ component of $g(\omega,\nu)$
gives a contribution to $\eta(\xi,r)$ of the form
\beq
\eta_{1}(\xi,r)= 2G_{1}\int_{0}^{\xi} d\xi'
\int d\nu f(\nu)E(\nu,r) \exp[\Delta(\nu)\xi]
=2G_{1}\xi \sigma(\xi,r)\, ,
\label{eq:9}
\endeq
which gives precisely the Regge growth of the diffraction slope
$B(\xi,r)$ with $\alpha_{\Pom}'=G_{1}$. We have an explicit
estimate for the slope of the pomeron trajectory
\beq
\alpha_{\Pom}' \sim
{3 \over 16\pi^{2}} \int d^{2}\vec{r}\,\,\alpha_{S}(r)
\mu_{G}^{2}r^{2}
K_{1}^{2}(\mu_{G}r)  \propto
{3\over 64\pi}R_{c}^{2}\alpha_{S}(R_{c})
\, .
\label{eq:10}
\endeq
The effect of $g_{2}(\omega,\nu)=G_{2}(\nu,\omega)$ can be evaluated
making use of the explicit form of $E(\nu,r)$ [3,4,7]. It also
contributes to the slope of the pomeron trajectory $\alpha_{\Pom}'
\sim G_{2}(0)\sim G_{1}$. The smooth part of $g(\omega,\nu)$ does
not contribute to the slope of the pomeron trajectory,

In the numerical calculation of the slope $\alpha_{\Pom}'$ we
start with the dipole-dipole cross section and the corresponding
diffraction slope as described in [5,6]. We calculate the
$\xi$ dependence of the dipole cross section $\sigma(\xi,r)$
and of the diffraction slope $B(\xi,r)$ and verify that at
$\xi \rightarrow \infty$ the effective intercept
$\Delta_{eff}(\xi,r) = \partial \log\sigma(\xi,r) /\partial\xi$
and the effective slope $\alpha_{eff}'(\xi,r)=\partial B(\xi,r)
/\partial \xi$ tend to the limiting
values $\Delta_{\Pom}$ and $\alpha_{\Pom}'$, respectively,
which are independent of the size of the projectile and target
colour dipoles. We take the running coupling with the infrared
freezing $\alpha_{S}(r) \leq \alpha_{S}^{(fr)}=0.82$ [3,4]. The
dependence of the
slope $\alpha_{\Pom}'$ on $\mu_{G}=1/R_{c}$ is shown in Fig.1.
The simple estimate (\ref{eq:10}) is close to these numerical
results.

To summarize, we have shown that the generalized BFKL pomeron
[1-4] is the moving cut in the complex angular momentum plane.
We derived a simple analytical estimate (\ref{eq:10}) for the
slope $\alpha_{\Pom}'$ of the pomeron trajectory and found the
dependence of the slope on the gluon correlation radius by an
accurate numerical solution of our generalized BFKL equation
(\ref{eq:2}) for the diffraction slope.

This work was supported by the INTAS Grant 93-239. V.R.Z.
acknowledged the partial support by the Grant NMT5000 of
the G.Soros ISF.
\pagebreak\\

{\bf Figure caption:\\}
{\bf Fig. 1} - The slope of the pomeron trajectory $\alpha_{\Pom}'$
as a function of the inverse correlation radius $\mu_{G}=1/R_{c}$
for the perturbative gluons.
\end{document}